\def\rn{\noindent\parshape 2 0truecm 8.5truecm 0.3truecm 8.2truecm}
\def\rn{}
\def\nn#1 #2{#2. #1}				
\def\nnn#1 #2 #3{#2. #3. #1}			
\def\nnnn#1 #2 #3 #4{#2. #3. #4 #1}		
\def\nnnnn#1 #2 #3 #4 #5{#2. #3. #4 #5. #1}	
\def\dualand{ and\hbox{ }}				
\def\multiand{, and\hbox{ }}				
\def\rf#1;#2;#3;#4;#5 {{\frenchspacing\par\rn#1, #3 {\bf #4}, #5 (#2). \par}}
\def\rg#1;#2;#3;#4;#5;#6 {{\frenchspacing\par\rn#1, #3 {\bf #4}, #5 (#2). \par}}
\def\rfbook#1;#2;#3;#4;#5 {{\frenchspacing\par\rn#1, {\it #3} (#5, #4, #2).\par}}
\def\rfprep#1;#2;#3 {{\par\frenchspacing\rn#1, #3 (#2).\par}}

\def\mK{{\rm \mu K}}

\def\expec#1{\langle#1\rangle}

\def\etal{{\frenchspacing\it et al.}}
\def\ie{{\frenchspacing\it i.e.}}
\def\eg{{\frenchspacing\it e.g.}}
\def\etc{{\frenchspacing\it etc.}}

\def\beq#1{\begin{equation}\label{#1}}
\def\eeq{\end{equation}}
\def\beqa#1{\begin{eqnarray}\label{#1}}
\def\eeqa{\end{eqnarray}}
\def\eq#1{equation~(\ref{#1})}

\def\eqn#1{~(\ref{#1})}

\def\fig#1{Figure~\ref{#1}}
\def\Fig#1{Figure~\ref{#1}}

\def\sec#1{Section~\ref{#1}}

\def\app#1{Appendix~\ref{#1}}

\def\spose#1{\hbox to 0pt{#1\hss}}
\def\simlt{\mathrel{\spose{\lower 3pt\hbox{$\mathchar"218$}}
     \raise 2.0pt\hbox{$\mathchar"13C$}}}
\def\simgt{\mathrel{\spose{\lower 3pt\hbox{$\mathchar"218$}}
     \raise 2.0pt\hbox{$\mathchar"13E$}}}
\def\simpropto{\mathrel{\spose{\lower 3pt\hbox{$\mathchar"218$}}
     \raise 2.0pt\hbox{$\propto$}}}

\def\ed{\end{document}}

\def\draft{
}

\def\arcdeg{^\circ}

\def\l{\ell}

\def\n{{\bf n}}
\def\r{{\bf r}}
\def\x{{\bf x}}
\def\y{{\bf y}}
\def\z{{\bf z}}

\def\xt{\tilde{\x}}
\def\xw{{\x}^w}

\def\bzero{{\bf 0}}
\font\bfmath=cmmib10
\def\err{\hbox{\bfmath\char'042}}	
\def\rh{\widehat{\r}}
\def\A{{\bf A}}

\def\I{{\bf I}}
\def\L{{\bf L}}

\def\N{{\bf N}}

\def\R{{\bf R}}
\def\SS{{\bf S}}

\def\Sig{{\bf\Sigma}}

\def\tr{\hbox{tr}\,}

\documentstyle[prd,aps,epsf]{revtex}
\draft
\begin{document}
\twocolumn[\hsize\textwidth\columnwidth\hsize\csname@twocolumnfalse\endcsname



\title{Comparing and combining the Saskatoon, QMAP and COBE CMB maps}

\author{Yongzhong Xu$^1$, Max Tegmark$^1$, Angelica de Oliveira-Costa$^1$,
Mark J. Devlin$^1$, Thomas Herbig$^2$, Amber D. Miller$^3$, 
C.~Barth Netterfield$^4$, Lyman Page$^2$}

\address{$^1$Dept. of Physics, Univ. of Pennsylvania, Philadelphia, PA 19104;
  xuyz@physics.upenn.edu}
\address{$^2$Princeton University, Dept. of Physics, Princeton, NJ 08544} 
\address{$^3$Dept. of Astronomy and Astrophysics, Univ. of Chicago, Chicago, IL 60637}
\address{$^4$Dept. of Physics and Astronomy, University of Toronto, Toronto, ON M5S1A7, Canada}

\date{Submitted to Phys. Rev. D Oct. 27 2000, revised Dec. 22}

\maketitle

\begin{abstract}
We present a method for comparing and combining maps with
different resolutions and beam shapes, and apply it to the
Saskatoon, QMAP and COBE/DMR data sets.
Although the Saskatoon and QMAP maps detect signal at
the $21\sigma$ and $40\sigma$ levels, respectively, their 
difference is consistent with pure noise, placing strong 
limits on possible systematic errors.
In particular, we obtain quantitative upper limits on relative 
calibration and pointing errors.
Splitting the combined data by frequency shows similar 
consistency between the Ka- and Q-bands, placing limits on
foreground contamination. The visual agreement between the maps
is equally striking.
Our combined QMAP+Saskatoon map, 
nicknamed QMASK, is 
publicly available at {\it www.hep.upenn.edu/$\sim$xuyz/qmask.html}
together with its $6495\times 6495$ noise covariance matrix.
This thoroughly tested data set covers a large enough area 
(648 square degrees --- currently the largest degree-scale map available)
to allow a statistical comparison with COBE/DMR, showing good agreement.
\end{abstract}

\pacs{98.62.Py, 98.65.Dx, 98.70.Vc, 98.80.Es}


] 


\begin{figure}[tb]
\vskip-1.2cm
\centerline{\epsfxsize=8.7cm\epsffile{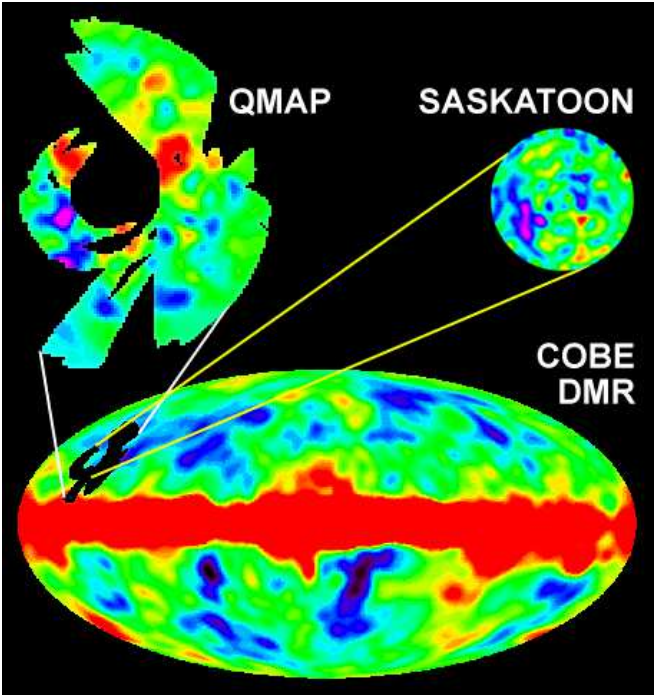}}
\caption{\label{AitoffFig}\footnotesize%
The three maps that we will compare and combine are shown in
Galactic coordinates. QMAP location in COBE map is shown in black.
}
\end{figure}

\section{INTRODUCTION}

The cosmic microwave background (CMB) field is currently enjoying
a bonanza of new high-quality data \cite{Gawiser00,deBernardis00,Hanany00},
which has triggered a surge of new papers about the implications for cosmological 
parameters
\cite{Lange00,boompa,Bambi00,Bridle00,observables,Jaffe00,Kinney00,concordance}.
There is currently such wide interest in these cosmological results that it is 
tempting to temporarily ignore underlying assumptions.
However, it is nonetheless important to consolidate these gains by careful 
study of the many technical analysis steps upon which they rest.
This can be done at many levels.
The experimental teams generally test for systematic
errors at many steps in their data analysis pipeline, from 
data acquisition, cleaning, calibration and pointing reconstruction
to mapmaking and power spectrum estimation.
In addition, numerous detailed comparisons have been made
between the angular power spectra $C_\l$ measured by different
experiments to determine whether they are all consistent
\cite{Scott95,Lineweaver98,9par,Dodel99,10par,Jaffe00}.
However, such comparisons use only a very small fraction
of the information at hand: average band powers, not the spatial
phase information.
A more powerful test (in the statistical sense of being more likely to
discover systematic errors) involves a direct comparison of the sky 
maps from experiments that overlap in both spatial and angular coverage.

Such a comparison is straightforward for maps with identical resolution and beam shape,
simply testing whether the difference map is consistent with pure detector noise.
Such tests have been successfully performed for the COBE/DMR maps 
\cite{Bennett96}.
Unfortunately, comparisons are usually complicated by
angular resolution differences between channels. 
Some experiments probe 
the sky in an even more complicated way, with, {\eg}, elliptical beams, 
double beams, triple beams, interferometric beams or complicated elongated 
software-modulated beams. Correlated noise further complicates
the problem.
Despite these difficulties, 
accurate comparisons between different experiments are 
crucial. Some of the best testimony to the quality of CMB maps
comes from the success of such comparisons
in the past --- between FIRS and DMR \cite{Ganga93},
Tenerife and DMR \cite{Lineweaver95},
MSAM and Saskatoon \cite{Netterfield97,Knox98},
two years of Python data \cite{Ruhl95},
three years of Saskatoon data \cite{saskmap},
two flights of 
MSAM \cite{Inman97}
and different channels of 
QMAP \cite{qmap1,qmap2,qmap3},
Boomerang \cite{deBernardis00}
and Maxima \cite{Hanany00}.

General methods have been developed for both comparing \cite{Knox98,comparing}
and combining \cite{comparing}
arbitrarily complicated experiments.
In this paper we will derive a technique for simplifying this task in practice,
and apply it to compare and combine the three overlapping data sets shown in 
\fig{AitoffFig}: COBE/DMR, Saskatoon and QMAP.
Our motivation is threefold:
\begin{itemize}
\item To test and provide methods that can be used by experimental groups
      in the future.
\item To search for systematic problems that may be relevant to 
      ongoing and future experiments.
\item To quality-test and publicly release the largest degree-scale map
      to date.
\end{itemize}
We stress that combining maps is not just a matter of
making pretty pictures. Power spectra from different experiments are
routinely combined as if their sample variance were independent.
However, since this approximation breaks down whenever the underlying maps overlap 
in spatial and angular coverage, the only correct way to compute their
combined power spectrum is to extract it from the combined map.

We present our methods in \sec{METHODSec}, present our results in \sec{ResultsSec}
and summarize our conclusions in \sec{ConclusionsSec}.


\section{METHOD}

\label{METHODSec}

We use the methods for combining and comparing maps presented in 
\cite{comparing}, which are 
most easily expressed with matrix notation.
Given two data sets represented by the vectors $\y_1$ and $\y_2$, we write
\beq{vEq}
\y_1=\A_1\x+\n_1, \quad\y_2=\A_2\x+\n_2.
\eeq
Here the vector $\x$ contains the temperature of the true sky at various
locations (pixels).
$\A_1$ and $\A_2$ are two known matrices incorporating the pointing strategy
and beam shape of each experiment.
$\bf n_1$ and $\bf n_2$ are two
random noise vectors with zero mean and known covariance matrices
$\N_1\equiv\expec{\n_1\n_1^t}$ and 
$\N_2\equiv\expec{\n_2\n_2^t}$.
It is convenient to define larger matrices and vectors
\beq{MatrixEq} 
\A\equiv\left(\begin{array}c
	\A_1 \\ \A_2 \end{array}\right), \quad
   \y \equiv \left(\begin{array}c
   	\y_1 \\ \y_2 \end{array}\right), \quad
   \n \equiv \left(\begin{array}c
   	\n_1 \\ \n_2 \end{array}\right)
\eeq
and to write the full noise covariance matrix as
\beq{covEq}
\bf N \equiv \langle nn^t\rangle= \left(\begin{array}{cc}
	\bf N_1 & \bf N_{12}\\
	\bf N_{12}^t & \bf N_2 \end{array}\right).
\eeq

We review the mathematical details of
combining, filtering and comparing maps in Appendices A, B and C, 
with some explicit details added (beyond \cite{comparing}) that are useful when
implementing these methods in practice.

We derive a new deconvolution method in \app{DeconvSec} which substantially
simplifies our calculations by eliminating the $\A$-matrices above.
In the generic case, deconvolution is strictly speaking impossible, 
since the matrix $\A$ is not 
invertible and certain pieces of 
information about $\x$ are simply not present in $\y$.
It is common practice to find approximate solutions to such under-determined
problems using singular value decomposition or other techniques, but
our goal is different: we need a deconvolved sky map $\xt$
that can be analyzed as a true sky map with $\A=\I$ without approximations,
shifting all complications into the new noise covariance matrix.
The method derived in \app{DeconvSec} is found to be stable numerically,
and can be used both for ``unsmoothing'' low-resolution maps and to 
deconvolve more complicated oscillatory beam patterns such as that of
the Saskatoon experiment.

\section{RESULTS}

\label{ResultsSec}
In this section, we combine the QMAP, Saskatoon and COBE maps. We then 
perform a battery of tests for systematic errors by comparing 
the maps with each other, paying particular attention to possible 
calibration, pointing and foreground problems.

\subsection{Saskatoon Data}

The Saskatoon data set is very different from other data sets such as QMAP and
COBE since it does not contain simple sky temperature measurements.
Instead, the 2970 Saskatoon measurements are
different linear combinations of the sky temperatures with rather complicated
weight functions reminiscent of caterpillars --- examples are plotted in 
\cite{Netterfield97,saskmap}.
These measurements probe a circular sky patch 
with about $16\arcdeg$ diameter, centered on the 
the North Celestial
Pole (NCP). 
In addition to the 2590 weight functions used in \cite{saskmap}, 
which are all oriented like spokes of a wheel,
we include the 380 ``RING'' data measurements, 
linear combinations in the perpendicular direction going around
the periphery of the observing region.

\begin{figure}[tb]
\vskip-1.7cm
\centerline{\epsfxsize=8.5cm\epsffile{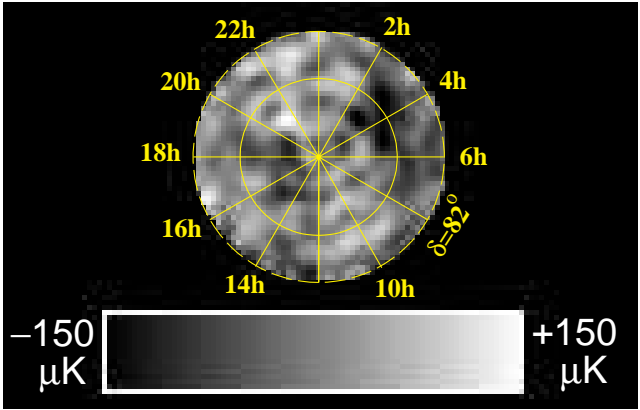}}
\vskip-1.85cm
\smallskip
\caption{\label{SaskFig}\footnotesize%
Wiener-filtered Saskatoon map. 
The CMB temperature is shown in
coordinates where the north celestial pole is at the center of a circle of
$16\arcdeg$ diameter, with R.A. being zero at the top and increasing clockwise.
In addition to the data included in the map of
\protect\cite{saskmap},
``RING'' data is included here.
Note that the orientation of this and all following maps is different
from that in \fig{AitoffFig}.
}
\end{figure}

We pixelize this sky region into 2016 pixels in the same coordinate
system as QMAP, \ie, a simple square grid in gnomonic equal area projection,
and define $\x$ to be the true sky convolved with a Gaussian beam of 
FWHM $0.68^\circ$.
We compute the $2970\times 2016$ matrix $\A$ using the software from the
original Saskatoon analysis \cite{Netterfield,Netterfield97}. The rows of $\A$
have a vanishing sum since the beam functions are all insensitive
to the monopole --- they are normalized so that the absolute values 
sum to two.
Using equations\eqn{newsimEq} and\eqn{newsimSig}, we obtain a deconvolved
Saskatoon CMB map $\xt_{\rm SASK}$ and its corresponding 
covariance matrix $\Sig_{\rm SASK}$. \Fig{SaskFig} shows the
Wiener-filtered Saskatoon map, which is visually almost identical 
to that in \cite{saskmap} except for the additional information
from the RING data near the border.

\begin{figure}[tb]
\centerline{\epsfxsize=8.5cm\epsffile{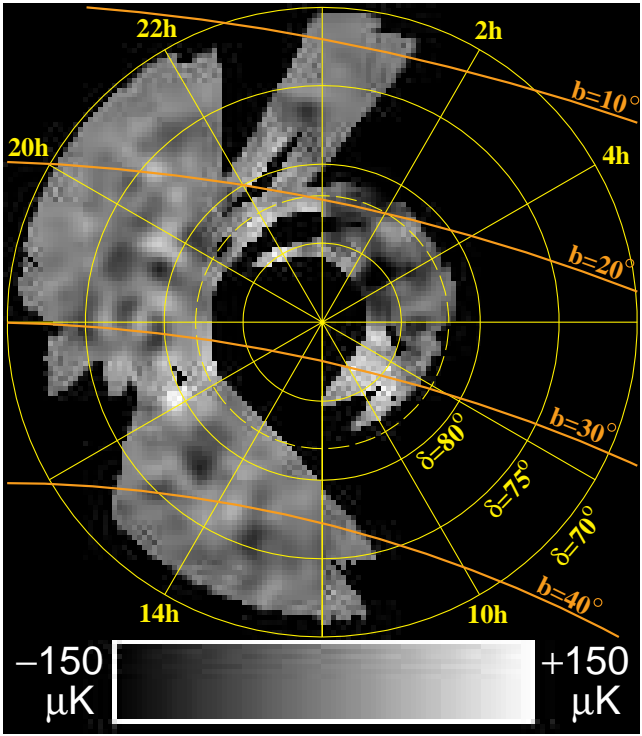}}
\smallskip
\caption{\label{QMAPfig}\footnotesize%
Wiener-filtered QMAP map. 
The coordinates are the same as in the previous figure.
}
\end{figure}

\begin{figure}[tb]
\centerline{\epsfxsize=8.5cm\epsffile{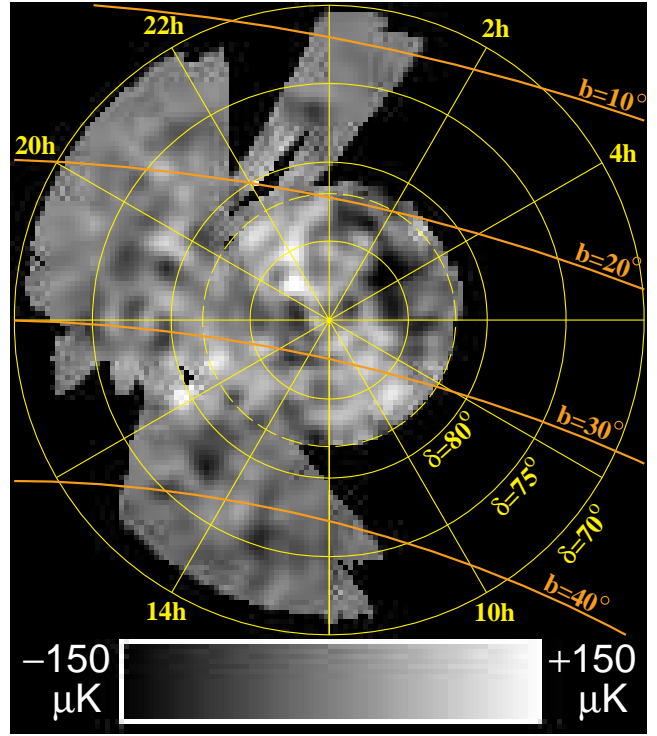}}
\smallskip
\caption{\label{qsaskFig}\footnotesize%
Wiener-filtered map combining the QMAP and Saskatoon experiments. The coordinates
are the same as in \fig{SaskFig}.
}
\end{figure}

\begin{figure}[tb]
\centerline{\epsfxsize=8.5cm\epsffile{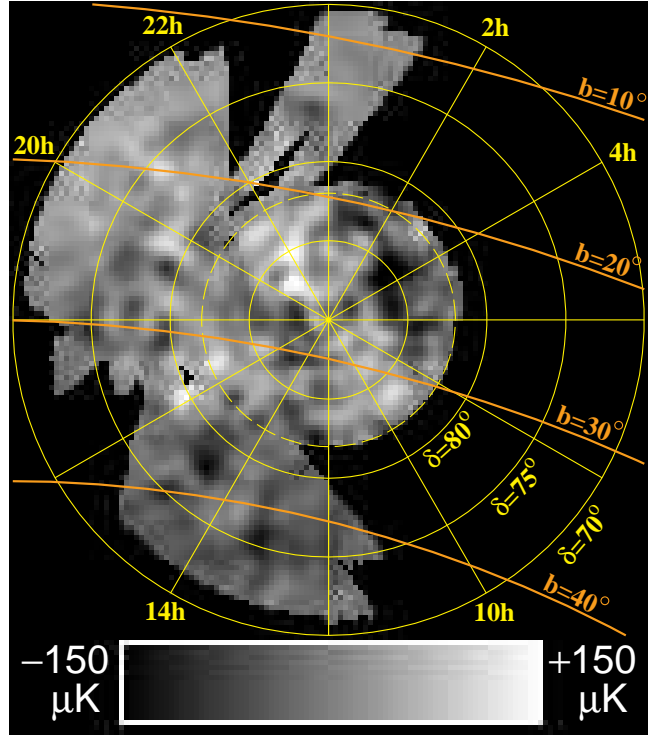}}
\smallskip
\caption{\label{qsaskcobeFig}\footnotesize%
Wiener-filtered map of the combined QMASK and COBE data. The coordinates
are the same as in the previous figure. COBE adds only large-scale information.
For example, the upper left region is brightened somewhat.
}
\end{figure}

\subsection{Combining QMAP with Saskatoon}
\label{ComboSec}

The QMAP data consists of Ka-band and Q-band measurements 
with angular resolution $0.89\arcdeg$
and $0.68\arcdeg$, respectively. We first deconvolve the Ka-band 
data to $0.68\arcdeg$ resolution using the method of \app{DeconvSec}.
We then produce a unified QMAP data set 
$(\xt_{\rm QMAP},\Sig_{QMAP})$
with a single resolution,
$0.68\arcdeg$, by combining the result with 
the Q-band as described in 
\app{CombinSec}, using \eq{FinalSig} to compute the 
final noise covariance $\Sig_{QMAP}$ since the overlap between
the two bands is only partial.
This combined QMAP map has 5396 pixels covering a sky area of about 538
square degrees.

Combining QMAP and SASK
is now a straightforward task, since the data sets $\xt_{\rm QMAP}$ 
and $\xt_{\rm SASK}$ have the same angular resolution and pixelization scheme. 
Since the spatial overlap is only partial, we once again use 
\eq{FinalSig} to compute the 
combined noise covariance matrix. 
There are 917 overlapping pixels, so the combined map consists of 6495 pixels,
covering a sky area of about 648 square degrees. We will nickname the
combined data set ``QMASK''. 

The main improvement in the combined is not the area covered (the QMASK map 
is only 20\% larger than QMAP), but the signal-to-noise and the topology.
SASK has excellent signal-to-noise in the region 
that it covers, which overlaps the most sensitive region of QMAP.
Indeed, the two maps have comparable sensitivity in the overlap region,
so both of them have substantial impact on the spatial features 
seen in Fig. 3.
Filling in the ``hole'' in the map is also useful for comparing
with lower resolution maps like COBE and for potential future applications, 
\eg, genus statistics, where 
a large contiguous area is desirable.

\subsection{Combining QMASK with COBE}

The COBE data \cite{Smoot,Bennett}
has much lower angular resolution than QMASK
(about $7.08\arcdeg$), and the pixel
size of COBE is much bigger than that of QMASK as well
(about $2.6\arcdeg\times 2.6\arcdeg$). 
In total, there are 6144
pixels in the whole COBE sky map. We select those pixels whose
centers are within the QMASK map and at least $3\arcdeg$ away from the 
perimeter. 
Only 58 COBE pixels satisfy these criteria. 

We first deconvolve this COBE data to the QMASK angular resolution, 
using the method of \app{DeconvSec} with
$\bf A_1$ being a $58\times 6495$ matrix with a Gaussian COBE beam on each row,
normalized to sum to unity. 
As input, we use the inverse-variance weighted average of the 53 and 90 GHz COBE/DMR channels 
\cite{Bennett96}.
By construction, our resulting COBE and QMASK maps overlap each other perfectly,
so we obtain our combined map by simply using equations\eqn{sim2Eq} 
and\eqn{sim2Sig}.

At first glance, \fig{qsaskFig}
and \fig{qsaskcobeFig} look very similar. 
However, inspecting them more carefully reveals that
although the small scale patterns are the same, 
the upper part in \fig{qsaskcobeFig} is brighter than that in
\fig{qsaskFig} (this is related to the QMASK cold spot 
that we will discuss subsection~\ref{CompqsaskcobeSub}).
In other words, since COBE contains only large scale
information, this is precisely what it has added in 
\fig{qsaskcobeFig}, leaving the small scale structure 
unaffected.

\subsection{Comparing QMAP with Saskatoon}

As mentioned above, there 
are 917 pixels overlapping between the QMAP and
Saskatoon data sets. After extracting these pixels and their two noise covariance
matrices from the full maps, we can compare the two experiments using 
the null-buster test described in \app{CompaSec}.

\begin{figure}[tb]
\centerline{\epsfxsize=9cm\epsffile{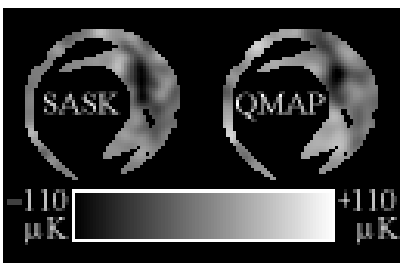}}
\end{figure}

\begin{figure}[tb]
\vskip-3.0cm
\centerline{\epsfxsize=9cm\epsffile{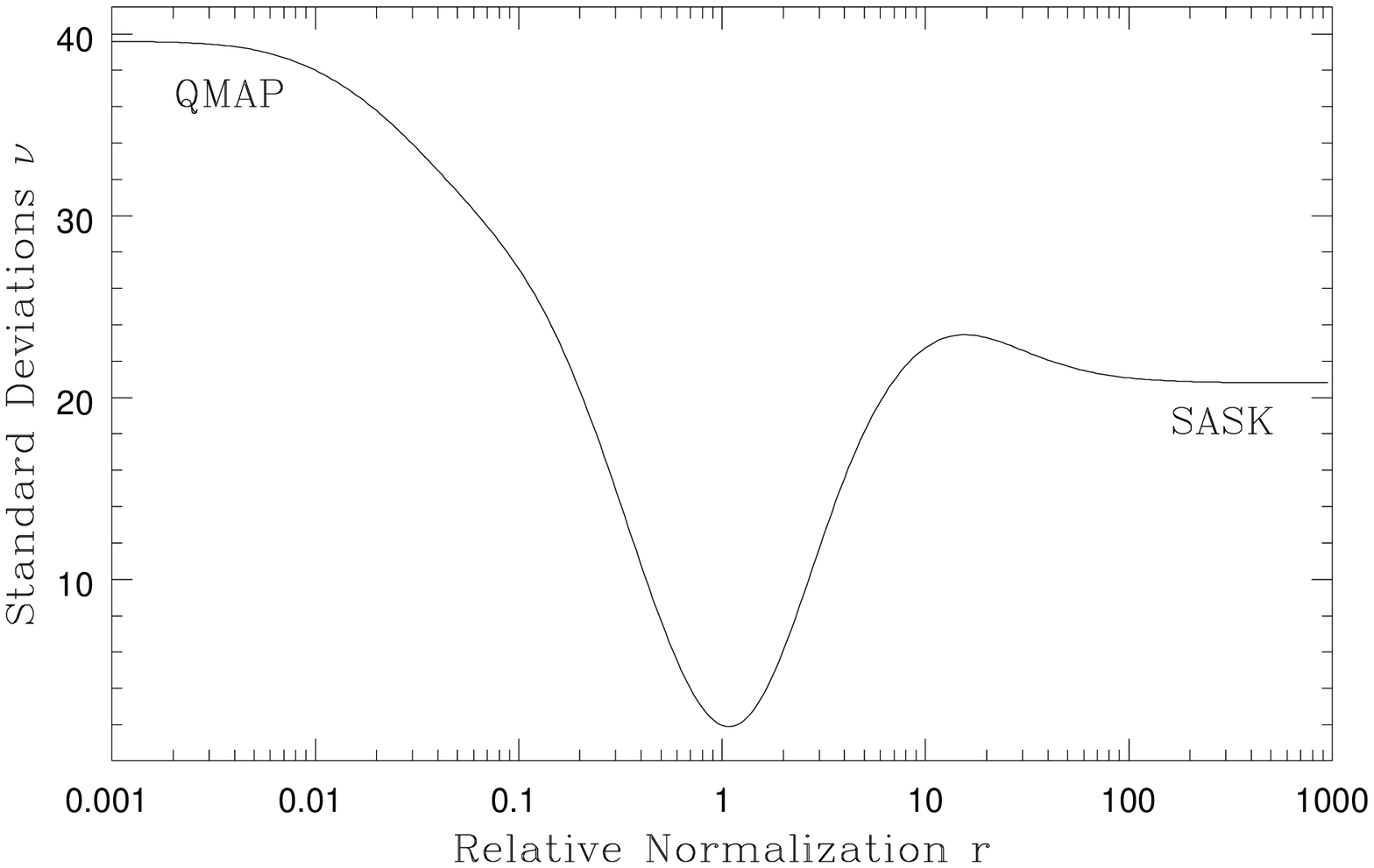}}
\caption{\label{qsaskSlopeFig}\footnotesize%
Comparison of QMAP (left) and Saskatoon (right). 
The upper panel shows both maps Wiener filtered with the same
weighting in the overlap region.
The lower panel shows the number of standard deviations 
(``sigmas'') at which the difference map 
$\xt_{ \rm QMAP}-r\xt_{\rm SASK}$ is inconsistent with mere noise.
Note that this is only for the overlap region.
}
\end{figure}

\subsubsection{Visual comparison}
\label{CompQMASKsubsub}

Before delving into statistical details, it is useful to compare the two maps 
visually. Comparing plots of the raw maps $\xt$ is rather useless, since they are
so noisy. Unfortunately, comparing plots of the two Wiener-filtered maps 
like figures~\ref{SaskFig}-\ref{qsaskcobeFig}
is not ideal either: since the noise matrices $\N_1$ and $\N_2$ are different,
this would entail comparing apples and oranges, since the two 
Wiener-filtered maps would be smoothed and weighted differently.
For instance, if a prominent spot in one map is invisible in the other,
this could either signal a problem or be due to that particular region 
being very noisy in the second map and therefore suppressed by the 
Wiener filtering.

To circumvent this problem, we Wiener-filter both maps exactly in the
same way, 
\beq{SpecialWienerEq}
\xw_i=\SS[\SS+\N_1+\N_2]^{-1}\xt_i,\quad i=1,2.
\eeq
In other words, we replace the individual noise covariance matrices
by their sum, so that the map will only show information that is 
accurately measured by {\it both} experiments.
These maps $\xw_1$ and $\xw_2$ are compared in the 
upper part of \fig{qsaskSlopeFig}, and are seen to 
look encouragingly similar.

\begin{figure}[tb]
\vskip-2.5cm
\centerline{\epsfxsize=9cm\epsffile{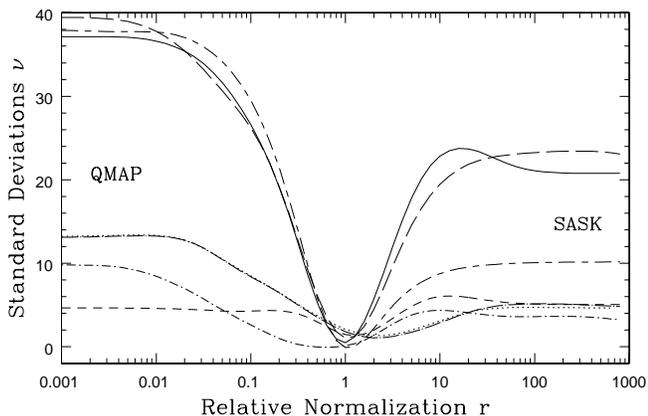}}
\caption{\label{AllSlopeFig}\footnotesize%
Same as previous figure, but comparing Saskatoon with various subsets of
the QMAP data. The curves correspond to 
Ka-band (dotted), Q-band (solid), flight 1 (short-dashed), flight 2 (long-dashed), 
f2Ka12(dot-long-dashed), f2Q12 (dot-short-dashed) and f2Q34 (short-dashed-long dashed).
}
\end{figure}

\subsubsection{Tests for systematic and calibration errors}

The lower plot in \fig{qsaskSlopeFig} shows the results of applying the null-buster test
to the difference map
$\xt_{ \rm QMAP}-r\xt_{\rm SASK}$ for different values of the constant $r$.
(The corresponding noise covariance matrix $\N=\N_1+\r^2\N_2$.)
The left part of the 
curve where $r\ll 1$ is dominated by information from QMAP, and we see
that QMAP alone (the $r=0$ case) is inconsistent with noise at 
about the $40\sigma$-level.
Similarly, we see that the Saskatoon map alone
(the case $\r=\infty)$ is inconsistent with noise at about
the $20\sigma$-level. Note that these significance levels are still higher
for the full maps --- here we are limiting ourselves to the sky region 
where they overlap.

In summary, both the QMAP and Saskatoon maps contain plenty of signal. 
Is this signal consistent between the two maps?
The answer is given by the most interesting point on the curve,
where the relative normalization value $r=1$.
In the absence of systematic or calibration errors, the corresponding difference map 
should contain pure noise. In our case, when $r=1$, the difference map 
$\xt_{\rm QMAP}-\xt_{\rm SASK}$ is seen to be consistent with pure noise, 
\ie, less than $2\sigma$ away from zero.
The strong signal seen in both maps therefore appears to be a true sky signal,
with no evidence for significant systematic errors in either QMAP or Saskatoon.

The QMAP experiment consists of two flights, each with two frequencies (Ka and Q-band)
and three slightly different
observing regions. We label these six sub-maps f1Ka12, f1Q2 and
f1Q34 (from flight 1) and f2Ka12, f2Q12 and f2Q34 (from flight 2). 
The pointing and calibration analyses for these two flights were 
completely separate, and the map-making algorithm was applied 
separately for these six sub-maps 
\cite{qmap1,qmap2,qmap3}.
To investigate possible problems with these individual sub-maps that may have
been averaged away in the combined analysis, we repeat the 
comparison with Saskatoon separately for each one. 
The results are shown in \fig{AllSlopeFig}. None of these curves 
show any evidence for
systematic or calibration errors. The f2Q34 map is seen to contain
the strongest signal, inconsistent with noise at the $38\sigma$ level.
This is because f2Q34 contains a striking cold spot --- we will return to this in more
detail in subsection~\ref{coldspotSub}. 

\bigskip
\bigskip

{\footnotesize
\textbf{Table 1.} Summary of map comparisons.
The first three lines give the number of ``sigmas'' at which 
map 1, the difference map and map 2, respectively, are inconsistent with noise.
The remaining lines give the best fit value and limits on the relative calibration 
$r$, or, for the Ka vs. Q case, the spectral index $\beta$.

\noindent
\begin{center}
\begin{tabular}{| c | c | c| c|}
\hline
	   & QMAP vs SK & QMASK vs COBE & Ka vs Q band \\ \hline
$\nu (r=0)$ & 40 & 62 & 40 \\
$\nu (r=1)$ & 1.97 & -0.64 & -0.53 \\
$\nu$($r=\infty$) & 21 & 3 & 26\\ \hline
$ r_{min}(3\sigma)$ & 0.79 & 0.09 & $-2.73^{*}$ \\
$ r_{min}(2\sigma)$ & 0.95 & 0.13 & $-2.2^{*}$\\
$ r_{min}(1\sigma)$ &  & 0.2 & $-1.7^{*}$\\
$ r_{best}$ & 1.1 & 1.25 & $ 0.^{*}$\\
$ r_{max}(1\sigma)$ &  & 11.7 & $1.42^{*}$\\
$ r_{max}(2\sigma)$ & 1.2 & 20 & $1.9^{*}$\\
$ r_{max}(3\sigma)$ & 1.48 & 63 & $2.2^{*}$\\
\hline
\end{tabular}
\end{center}

\noindent
* values of the power spectral index $\beta$ in subsection~\ref{ForegoundSub}
}

\bigskip
\bigskip

The null-buster curves are interesting at more than just the points $r=0, 1$,
and $\infty$; 
the entire region near $r=1$ places limits on calibration errors.
A relative calibration error of say 10\% would shift the minimum of the
curve sideways to 0.9 or 1.1, depending on whether the QMAP or Saskatoon map was too high.
We can therefore place limits on calibration errors by reading off the $r$-values 
where noise is ruled out at say 2 sigma.
For instance, the QMAP-Saskatoon comparison constrains the relative calibration
error to be less than 20\% ($3\sigma$), and the $r$-values for different
significance levels
are shown in table 1.
This method may prove quite useful for upcoming 
experiments that have higher sensitivity.

\subsubsection{Pointing tests}

Our comparison method can also be used to test for relative pointing errors,
as a complement to the standard lower-level pointing tests that are routinely made using 
point sources {\etc}
Although an overall sideways shift of a single map will not affect the measured
power spectrum, such errors can become disastrous if the map is combined with
another one. 

\begin{figure}[tb]
\vskip-0.5cm
\hglue0.5cm
\centerline{\epsfxsize=12cm\epsffile{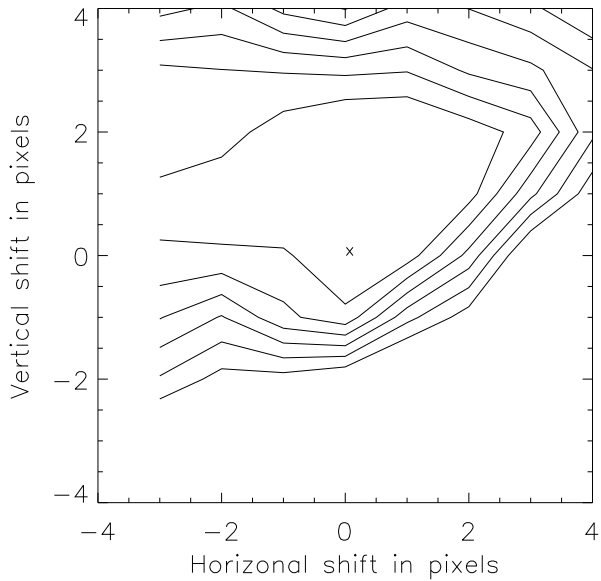}}
\caption{\label{ShiftFig}\footnotesize%
Test of the relative pointing of QMAP and Saskatoon.
The curves show the number of ``sigmas'' at which the difference map
is inconsistent with noise when the QMAP map is shifted 
vertically and horizontally. Starting from the inside, the contours are at 1,
2, 3, 4, 5, and $6\sigma$ respectively.
Cross indicates no shift.
The pixels are squares of side $0.3125^\circ$. 
}
\end{figure}

As an illustration of such a test, we compare the f2Q34 map 
with the Saskatoon map with the null-buster test (setting $r=1$)
after shifting it vertically and horizontally by an integer number of pixels.
\Fig{ShiftFig} shows that there is no
evidence for pointing error although we cannot give a strong constraint. 
Just as the calibration test, this pointing test based on CMB maps alone is likely to be
useful for upcoming high-sensitivity experiments. 

\begin{figure}[tb]
\centerline{\epsfxsize=9cm\epsffile{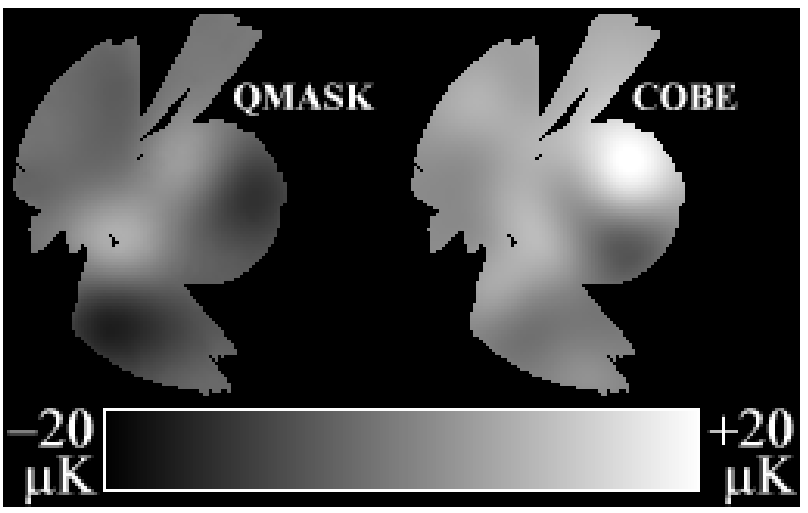}}
\end{figure}

\begin{figure}[tb]
\vskip-3.0cm
\centerline{\epsfxsize=9.0cm\epsffile{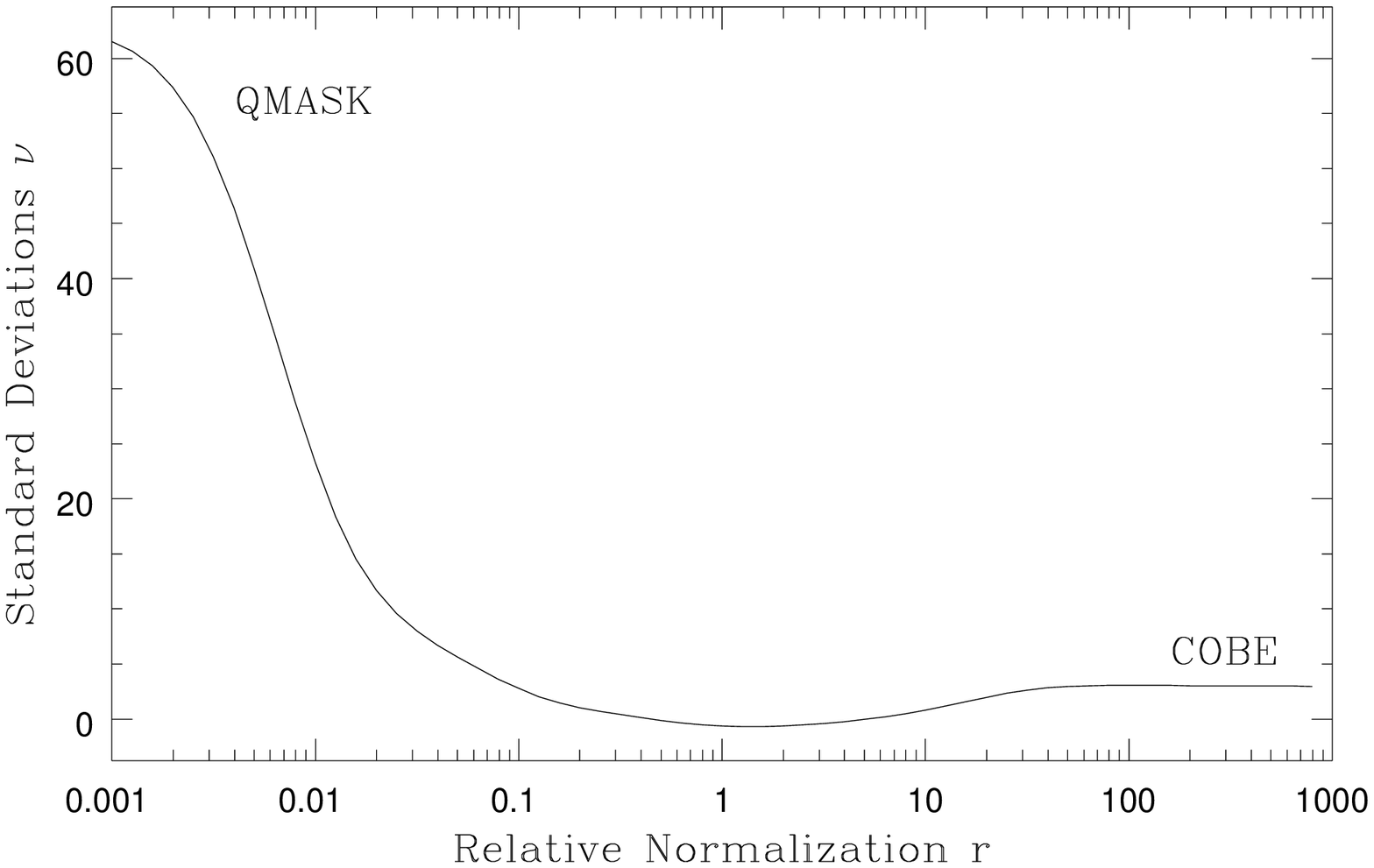}}
\caption{\label{qsaskcobeSlopeFig}\footnotesize%
Comparison of QMASK (left) with COBE (right). 
The upper panel shows both maps Wiener filtered with the same
weighting in the overlap region, using \eq{SpecialWienerEq}.
The lower panel shows the number of standard deviations 
(``sigmas'') at which the difference map 
$\xt_{ \rm QMASK}-r\xt_{\rm COBE}$ is inconsistent with mere noise,
and illustrates that the visual discrepancy at ``2 o'clock'' is
consistent with a fluctuation in the (correlated) noise.
}
\end{figure}

\subsection{Comparing QMASK with COBE}
\label{CompqsaskcobeSub}

Our QMASK map and our deconvolved COBE map cover the exact same sky region, 
so there are 6495 pixels in the overlap maps. 
Generating \fig{qsaskcobeSlopeFig}, which 
compares these two maps, therefore involved a marathon computer run,
processing the $6495\times 6495$ matrices of \eq{devEq} for each 
$r$-value.\footnote{
If CPU time had been an issue, this calculation could have been 
accelerated by binning the QMASK pixels
into larger ones. This would give essentially the same answer,
since the null-buster test gives statistical weight only to modes
where both maps are sensitive --- in this case, to large scale
modes only.
}
As can be seen in the lower panel of \fig{qsaskcobeSlopeFig}, the
QMASK data is inconsistent with noise at about the $62\sigma$-level,
whereas the COBE data is inconsistent with noise slightly above the 
$3\sigma$-level. Since neither the QMAP nor the Saskatoon experiments were
designed to probe such large angular scales, it is quite encouraging that
the QMASK-COBE difference map ($r=1$) is seen to be consistent
with pure noise.
Since the minimum of the curve is so broad, however, 
we obtain no interesting constraints on calibration errors.

The upper panel in \fig{qsaskcobeSlopeFig} shows that the two maps look 
fairly similar
considering the weak ($3\sigma$) COBE signal, 
with the notable exception of the upper right part of the Saskatoon disk.
Here COBE shows a hot spot whereas QMASK shows a cold spot.
We will return to this issue in more detail below, in 
subsection~\ref{coldspotSub}.
When generating these two maps, we used the same equal-weighting Wiener filtering
method that was described in 
subsubsection~\ref{CompQMASKsubsub}. This is particularly important here,
since the QMASK and COBE have such dramatically different angular resolutions
--- in contrast, a visual comparison of the Wiener filtered COBE map with
the normal Wiener filtered QMASK map from \fig{qsaskFig} is 
rather useless, since the latter is dominated by small-scale fluctuations.

\subsection{Foreground constraints}
\label{ForegoundSub}

The previous two sections used map comparisons to 
test for calibration and pointing errors.
Here we will compare maps at different frequencies to 
constrain the spectrum of the detected sky signal.

The presence of foreground contamination (see \cite{foregpars}
for a recent review) has been quantified for 
both the Saskatoon \cite{saskforeg} and QMAP \cite{qmapforeg} experiments
by cross-correlating the maps with various foreground templates.
The dominant foreground emission is expected to be due to 
synchrotron radiation, free-free emission and (vibrational and spinning) 
dust emission, from both the Milky Way (seen as diffuse emission) and
other galaxies (seen as point sources).
These cross-correlation analyses concluded that foregrounds played only a 
subdominant role in Saskatoon and QMAP.

By comparing the Ka- and Q-band maps, we are able to place a 
direct constraint on the frequency dependence of the signal.
Both Saskatoon and QMAP observe in both of these frequency bands.
We therefore repeat the analysis described above 
(Saskatoon deconvolution, merging with QMAP, \etc) separately for 
each of the two bands. The upper panel 
of \fig{KaQSlopeFig} shows equal-weighting 
Wiener-filtered maps for Ka-band and Q-band in the sky region 
that was observed at both frequencies, showing that they 
visually agree well. 

If we fit the frequency dependence by a power law
$\delta T(\nu)\propto\nu^{\beta}$ over the narrow frequency range in question
($\nu_{\rm Ka}\approx$ 30 GHz, 
 $\nu_{\rm Ka}\approx$ 40 GHz),
then characteristic spectral indices are
$\beta\sim -2.8$ for synchrotron, $\beta\sim -2.15$ for free-free emission,
$\beta\sim -3$ for spinning dust and 
$\beta\sim 2$ for vibrating dust.
By definition, $\beta=0$ for CMB.
If the sky signal in our maps obeyed $\delta T(\nu)\propto\nu^{\beta}$,
then the difference map $\xt_{\rm Ka}-r\xt_{\rm Q}$
would contain pure noise when $r$ was such that  
\beq{betaEq}
\beta={\lg r\over\lg(\nu_{\rm Ka}/\nu_{\rm Q})}. 
\eeq
The lower panel of \fig{KaQSlopeFig} is therefore plotted with
$\beta$ rather than $r$ on the horizontal axis.
Insisting that the difference map not be inconsistent with noise at more
than $1\sigma$ gives the spectral index constraint 
$\beta=0.0^{+1.4}_{-1.7}$, which is inconsistent with the signal being 
any single one of the foregrounds mentioned above.

\begin{figure}[tb]
\centerline{\epsfxsize=8.5cm\epsffile{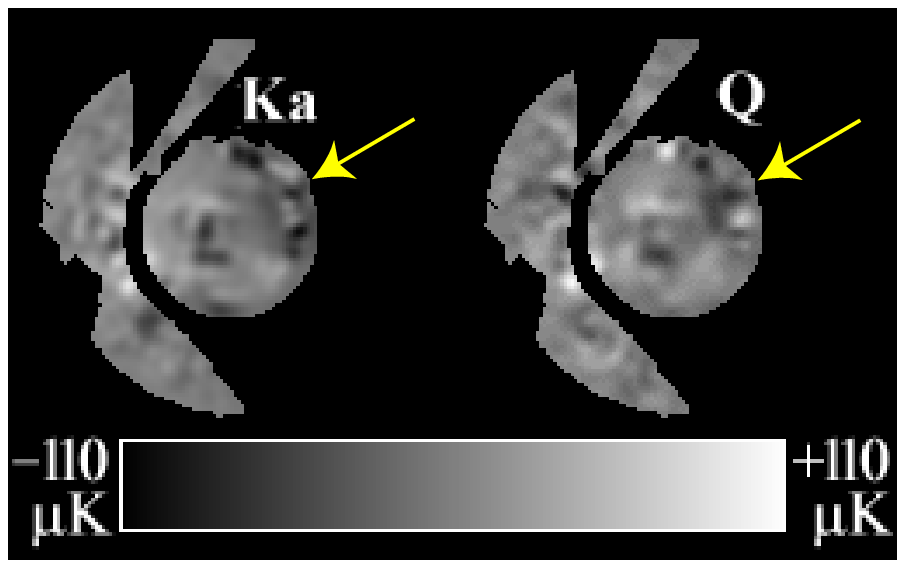}}
\end{figure}

\begin{figure}[tb]
\vskip-3.0cm
\centerline{\epsfxsize=9.0cm\epsffile{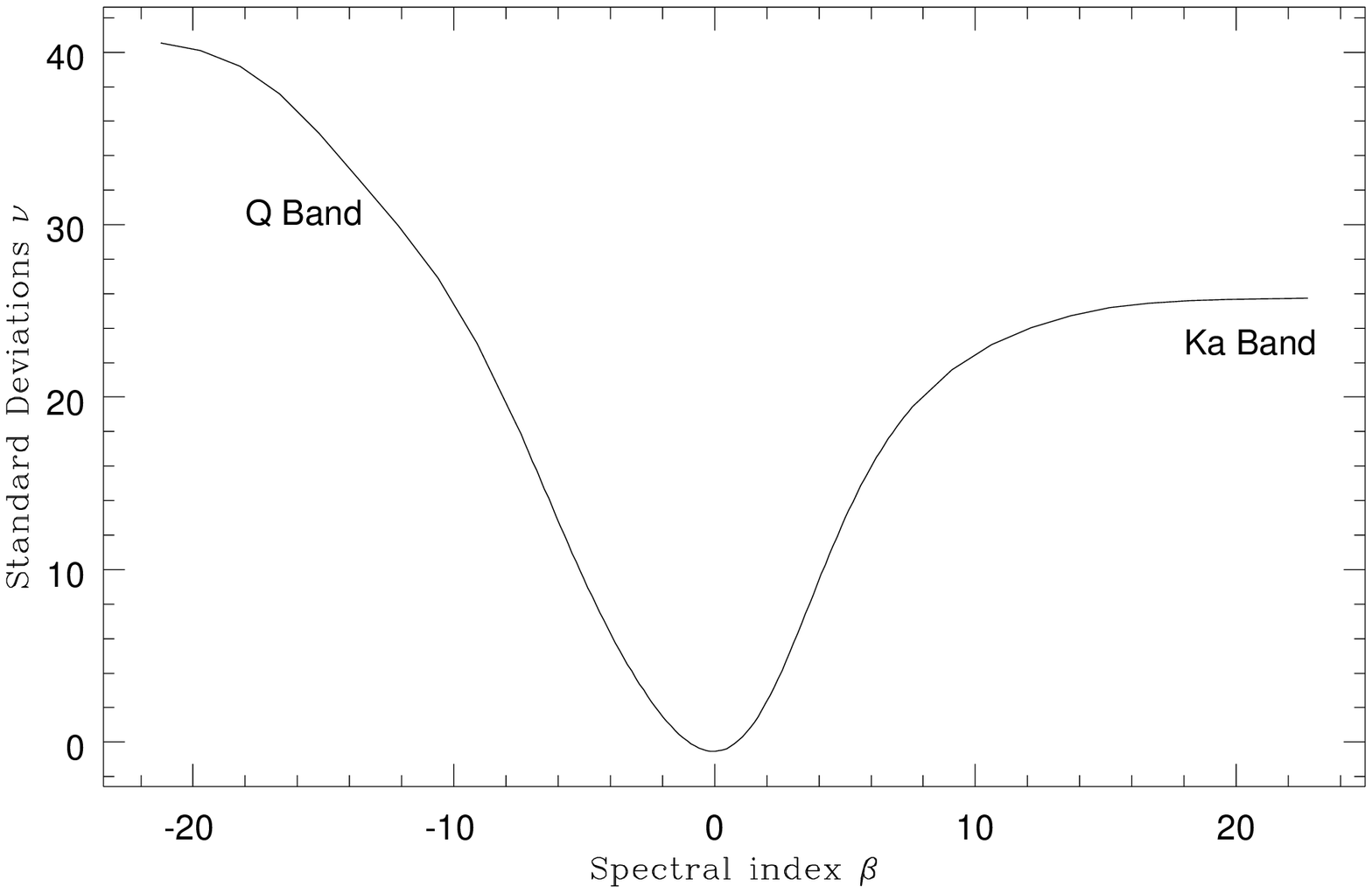}}
\caption{\label{KaQSlopeFig}\footnotesize%
Comparison of QMASK at two different frequencies, 
Ka-band (left) and Q-band (right).
The upper panel shows both maps Wiener filtered with the same
weighting in the overlap region that was observed at both frequencies.
Arrows indicate the coldest spot discussed below.
The lower panel shows the number of standard deviations 
(``sigmas'') at which the difference map 
$\xt_{ \rm Ka}-r\xt_{\rm Q}$ is inconsistent with mere noise.
}
\end{figure}

\subsection{The coldest spot}
\label{coldspotSub}

Up until now, we have presented a battery of tests for systematic errors
and other problems, all of which passed.
However, our maps did turn up one somewhat anomalous feature:
an unusually cold spot around ``two o'clock'' in the Saskatoon disk.
The spot's coldest pixel in the QMASK map is located at 
RA=$3^{h}20^{m}$, DEC=$84\arcdeg 55^{'}$.
Here the Wiener-filtered $Q$-band map in \fig{KaQSlopeFig} gives 
$\delta T\approx -230\mK$. For comparison, the expected rms fluctuations 
in this map are $27\mK$ from detector noise and $49\mK$ from
CMB fluctuations (for the ``concordance'' power spectrum of \cite{concordance}),
summing to $56\mK$ in quadrature. Taken at face value, this would indicate that
the spot is a $-4.1\sigma$ fluctuation.
For completeness, this section describes a number of additional tests
performed in an attempt to clarify its nature.

It is unlikely that the cold spot is due entirely to systematic error, since 
it is clearly detected by both QMAP (covered by the 
Q3 and Q4 detectors from in flight 2) and Saskatoon. The
Saskatoon experiment even detected this spot independently in each of its
three observing seasons \cite{Netterfield97,saskmap}.
Indeed, the reason that the f2Q34 map shows the most spectacular agreement
with Saskatoon in \fig{AllSlopeFig} is that f2Q34 is the only QMAP map that
covers the area containing this spot. 

\begin{figure}[tb]
\vskip-1.0cm
\centerline{\epsfxsize=9.0cm\epsffile{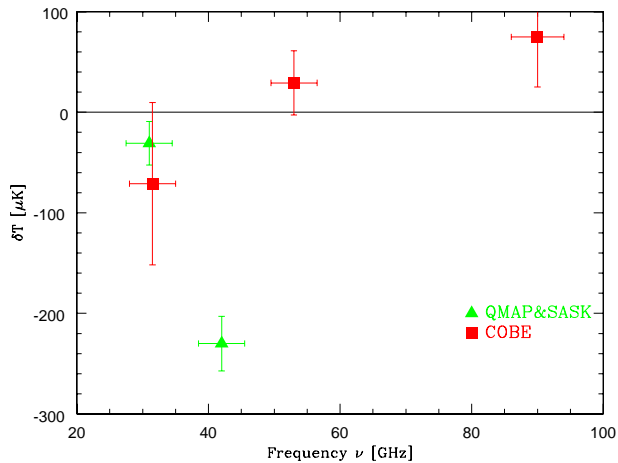}}
\vskip-1.5cm
\caption{\label{ColdestFig}\footnotesize%
The temperature towards
RA=$3^{h}20^{m}$, DEC=$84\arcdeg 55^{'}$
at different frequencies. Errors bars correspond to detector noise alone.
Note that these points cannot be interpreted as a spectrum of this sky
region, since the COBE points have much lower resolution 
and the two QMASK points have been Wiener filtered with different weights,
pushing them closer to zero than the underlying sky temperature.
}
\end{figure}

We have plotted all available microwave measurements of this region
in \fig{ColdestFig} as a function of frequency, including both
the Ka- and Q-band measurements from QMASK and the COBE/DMR
observations at 31.5, 53 and 90 GHz \cite{Bennett96}.
Unfortunately, \fig{ColdestFig} is 
not an actual spectrum
of the spot. It is more of a comparison of apples and oranges, 
since the measurements differ dramatically in angular resolution 
($7^\circ$ for COBE) and the QMASK maps are Wiener-filtered. 
Since Wiener filtering always pushes the signal towards zero
when noise is present, the QMASK points should be interpreted 
as lower limits on $|\delta T|$ --- the true sky temperature is likely
to be even colder. 

On the seven degree scale probed by COBE,
no evidence is seen for a cold spot (this can also be
seen in the map of \fig{qsaskcobeSlopeFig}), and the COBE spectrum of this
region appears consistent with CMB, \ie, $\beta=0$.
One possible interpretation is that a small cold region resolved by QMAP 
and Saskatoon is partly smoothed away by COBE.

Unfortunately, this sky patch is not covered by QMAP in Ka-band, so the Ka-information
comes from Saskatoon alone and is therefore noisier than the Q-band
measurement. This means that the Wiener-filtering has suppressed the Ka-band more,
as well as lowered its resolution by more aggressive smoothing.
However, the expected rms CMB fluctuations in the Wiener-filtered maps are
not nearly as different as the data points in \fig{ColdestFig}
($-31\mK$ at Ka-band and $-230\mK$ at Q-band), indicating that the 
low Q-band temperature does not persist fully down to Ka-band.
This argues against both a CMB origin and a
thermal SZ-origin, which would cause a cold spot that 
was essentially frequency-independent for $\nu<100$ GHz.

All other known microwave foregrounds produce hot rather than cold spots.
Some sort of absorption
process also appears unlikely, since the absorbing medium would have to be
colder than 3K.
No relevant foreground emission or X-ray cluster is found 
in radio, infrared or X-ray maps of the region.
Some adjacent dust emission is seen in the
IRAS 100$\mu$m map \cite{saskforeg}, which could potentially make 
this region look cold in contrast (since none of our maps are
sensitive to the monopole mode and measure merely relative temperatures), 
but only at the level of about $10\>\mu$K.

In Q-band, the spot is observed by QMAP only in flight two and only in the 
Q3 and Q4 channels. The latter dominates statistically, and has a 
13\% calibration uncertainty. However, Saskatoon also observes the spot 
in Q-band, and both experiments measure around $-200\mu$K in their individual 
Wiener filtered maps.

The explanation is fairly likely to be something mundane, since 
a 4-sigma fluctuation (which should happen about once for every
16,000 independent pixels) is not extremely unlikely when many pixels 
are considered --- if the effective number of independent regions in 
the Wiener filtered map is $10^2$ taking the smoothing into account,
the significance level gets downgraded to
$(1-1/16000)^{100}\approx 99\%$ \cite{gaussianity}.

In summary, the cold spot is definitely out there at some level,
but we have no single simple interpretation of what is causing it.
Its unusual spectrum argues against a non-Gaussian CMB fluctuation, 
foreground contamination and an SZ signal.
The most likely remaining explanation is a confluence of 
a less extreme CMB cold spot, noise fluctuations, calibration uncertainty and 
perhaps some small systematic error.
We have described this spot in such detail simply to ensure that no 
hints of problems with the data 
get swept under the rug.

The MAP satellite should resolve this puzzle next 
year, observing the spot with high sensitivity and resolution at 
22, 30, 40, 60 and 90 GHz.

\section{DISCUSSION}
\label{ConclusionsSec}

We have presented methods for comparing and combining CMB maps and applied
them to the QMAP, Saskatoon and COBE DMR data sets. 
We found that these methods were able to place interesting 
constraints on calibration and pointing problems, foreground
contaminations and systematic errors in general. This should make
them quite useful for ongoing and upcoming high-precision experiments. 

The data sets passed our entire battery of consistency tests,
placing strong limits on systematic errors.
Although the Saskatoon and QMAP maps detect signal at
the $21\sigma$ and $40\sigma$ levels in the overlap region, respectively, their 
difference is consistent with pure noise.
Our combined QMAP + Saskatoon map, 
nicknamed QMASK, covers a large enough area
to allow a statistical comparison with COBE/DMR, showing good agreement.

The one surprise that our battery of tests turned up is 
that a small region around (RA=$3^{h}20^{m}$,DEC=$84\arcdeg 55^{'}$)
appears unusually cold, mainly in Q-band.
Its unusual frequency dependence argues against non-Gaussian CMB fluctuations,
SZ-signal and known foregrounds.

The QMASK map presented here has been made 
publicly available at {\it www.hep.upenn.edu/$\sim$xuyz/qmask.html}
together with its $6495\times 6495$ noise covariance matrix.
With its 648 square degrees, this thoroughly tested data set 
is currently the largest degree-scale map available,
detecting signal at a level exceeding $60\sigma$.


\bigskip
Support for this work was provided by
NSF grants AST00-71213 and NSF 9732960,
NASA grant NAG5-9194 ({\it n\'e} NAG5-6034),
a NASA GSRP grant,
the University of Pennsylvania Research Foundation,
the Zaccheus Daniel Foundation and the Dodds Foundation.
The QMAP and SASK experiments were supported by NSF grants
PH 89-21378, 92-22952, 96-00015, NYI (to LP); NASA grants NAGW-1482,
NAGW-2801, NAG5-6034; Hubble Fellowships
HF-01044.01-93A (to TH) and HF-01084.01-96A (to MT); a Cottrell Award
from the Research Corporation (to LP) and a David and Lucile Packard
Foundation Fellowship (to LP).
The COBE data sets were developed by the NASA
Goddard Space Flight Center under the guidance of the COBE Science Working 
Group and were provided by the NSSDC.

\appendix

\section{Combining maps}
\label{CombinSec}
The combined map $\xt$ defined by
\beq{comvEq}
\xt\equiv\left[\A^t\N^{-1}\A\right]^{-1}\A^t\N^{-1}\y
\eeq
can be shown to be unbiased $(\expec{\xt}=\x)$, to minimize
the rms noise in each pixel and, 
if the noise properties are Gaussian, to retain all 
information about the true sky $\x$ that was present in the two original
maps \cite{comparing}.
The corresponding covariance matrix of the noise $\bf\epsilon\equiv\xt-\x$ is
\beq{SigcovEq}
\Sig\equiv\expec{\err\err^t} = 
\left[\A^t\N^{-1}\A\right]^{-1}.
\eeq
In all cases treated in this paper, the noise is uncorrelated between 
different maps (($\N_{12}=0$), which simplifies these equations to
\beqa{simEq}
 \xt&=&\Sig\left[\A_1^t\N_1^{-1}\y_1 + \A_2^t \N_2^{-1}\y_2\right],\\
 \label{simSig}
 \Sig&=&\left[\A_1^t\N_1^{-1}\A_1 + \A_2^t\N_2^{-1}\A_2\right]^{-1}.
\eeqa
Thanks to the deconvolution technique that will be described in 
\app{DeconvSec}, we will generally face the much simpler case
where the two data sets are two
sky maps with the exact same angular resolutions, 
\ie, the case $\A_1=\A_2=\I$, reducing the last two equations to simply
\beqa{sim2Eq}
  \xt&=&\Sig\left[\N_1^{-1}\y_1+\N_2^{-1}\y_2\right],\\
  \label{sim2Sig}
  \Sig&=&\left[\N_1^{-1} + \N_2^{-1}\right]^{-1}.
\eeqa
For the case of only a single pixel, we recognize this as a familiar
inverse-variance weighting. More generally, if the two noise matrices
can be simultaneously diagonalized, we see that this combination scheme
corresponds to an inverse-variance weighting eigenmode by eigenmode.

Generally maps overlap only partially, so we need only apply
this matrix method in the common region. Yet care needs to be 
taken in computing the noise covariance matrix $\Sig$ of the final 
map, since it will contain correlations between the common 
region and the rest.
Let us split the noise vectors for the two maps as
\beq{NewNoiseVec}
\n_1=\left(\begin{array}c \n_a \\ \n_{c1} \end{array}\right),\quad
\n_2=\left(\begin{array}c \n_b \\ \n_{c2} \end{array}\right),
\eeq
where the subscript $c$ refers to the common region of the two maps
whereas $a$ and $b$ refer to the regions that only belong to
maps 1 and 2, respectively.
We write the corresponding covariance
matrices as
\beq{NewCovMat}
\N_1\equiv\expec{\n_1\n_1^t} = \left(
   \begin{array}{cc}
   \N_a & \expec{\n_a\n_{c1}^t}\\
   \expec{\n_{c1}\n_a^{t}}& \N_{c1}
   \end{array}
   \right),
\eeq
\beq{NewCovMat2}
\N_2\equiv\expec{\n_2\n_2^t} = \left(
   \begin{array}{cc}
   \N_b & \expec{\n_b\n_{c2}^t}\\
   \expec{\n_{c2}\n_b^{t}}& \N_{c2}
   \end{array}
   \right),
\eeq
Substituting \eq{vEq} into \eq{simEq},
we obtain the noise vector $\n_c$ for the combined map in the common
region:
\beq{ComNoiseVec}
\n_c=\Sig_c\left[\A_{c1}^t\N_{c1}^{-1}\n_{c1}
		+\A_{c2}^t\N_{c2}^{-1}\n_{c2}\right],
\eeq
where $\Sig _c$ is given by \eq{simSig} for the common part. 
The final combined noise
covariance matrix $\Sig$ for the 
noise vector $(\n_a,\n_b,\n_c)$ of the combined map 
is therefore
\beq{FinalSig}
\Sig =\left(\begin{array}{ccc}
   \N_a & \bzero & \expec{\n_a\n_c^t}\\
   \bzero & \N_b & \expec{\n_b\n_c^t}\\
   \expec{\left[\n_a^o\n_c^t\right]^t}&
   \expec{\left[\n_b\n_c^t\right]^t}& \Sig_c
   \end{array}
   \right),
\eeq
where 
\beqa{CrossCorrEq}
\expec{\n_a\n_c^t} &=& \expec{\n_a\n_{c1}^t}\N_{c1}^{-1}\A_{c1}\Sig_c,\\
\expec{\n_b\n_c^t} &=& \expec{\n_b\n_{c2}^t}\N_{c2}^{-1}\A_{c2}\Sig_c.
\eeqa  
We need to use these expressions repeatedly in this paper to combine
partially overlapping maps, \eg, combining the different QMAP
flights with each other and combining QMAP with Saskatoon.

\section{Plotting maps}

Although the map $\xt$ contains all the sky information from an experiment,
plotting it is not very useful when some modes are much more noisy than others, 
thereby dominating the visual image. For this reason, it has become 
standard in the community to plot the corresponding Wiener filtered map,
defined as
\beq{WienerEq}
\x_w\equiv\SS[\SS+\N]^{-1}\xt,
\eeq
where $\SS$ is an estimate of the covariance matrix due to sky signal.
Throughout this paper, we use the $\SS$-matrix corresponding 
to the ``concordance'' power spectrum from \cite{concordance},
which agrees well with all current CMB measurements.

\section{Comparing maps}

\label{CompaSec}

Here we discuss the issue of how to test whether two data sets are consistent
or display evidence of systematic errors. Specifically, is there 
some true sky $\x$ such that the data sets $\y_1$ and $\y_2$ 
are consistent with \eq{vEq}?
Let us first consider the simplest case where the two data sets
sample the sky in the same way, that is, 
$\A_1=\A_2$. 
Consider two hypotheses: 
\begin{itemize}
\item[$H_0$:]The null hypothesis $H_0$ that there are no systematic errors,
so that the difference map $z\equiv\y_1-\y_2$ consists of pure noise
with zero mean and covariance matrix $\expec{\z\z^t}=\N\equiv\N_1+\N_2$.
\item[$H_1$:]The alternative hypothesis that the difference map
$\z$ consists of some signal besides noise, \ie, 
$\expec{\z}=0$, and $\expec{\z\z^t}=\N+\SS$ for some signal covariance
matrix $\SS$. 
\end{itemize}
The  ``null-buster'' statistic \cite{comparing}
\beq{devEq}
  \nu \equiv \frac{\z^t\N^{-1}\SS\N^{-1}\z - \tr\{\N^{-1}\SS\}}
  {\left[2\>\tr\{\N^{-1}\SS\N^{-1}\SS\}\right]^{1/2}}
\eeq
can be shown to rule out the null hypothesis $H_0$ with the largest average
significance $\expec{\nu}$ if $H_1$ is true, and can be interpreted as
the number of ``sigmas'' at which $H_0$ is ruled out \cite{comparing}.
Note that for the special case $\SS\propto\N$, it reduces to simply
$\nu=(\chi^2-n)/\sqrt{2n}$, where 
$\chi^2\equiv\z^t\N^{-1}\z$ is a standard chi-squared statistic.
The null-buster test can therefore be viewed as a generalized 
$\chi^2$-test which places more weight on those particular 
modes where the expected signal-to-noise is high.
It has proven successful comparing both 
microwave background maps \cite{qmap1,qmap2,qmap3} and  
galaxy distribution \cite{r,EfstathiouNullbuster}.

To evaluate \eq{devEq} in practice, it is useful to
Cholesky decompose the noise matrix as $\N=\L\L^t$ and compute the matrix
$\R\equiv\L^{-1}\SS\L^{-t}$. The remainder of the calculation now becomes 
trivial, since $\tr\{\N^{-1}\SS\}=\tr\R=\sum\R_{ii}$ and 
$\tr\{\N^{-1}\SS\N^{-1}\SS\}=\tr\R^2=\sum(\R_{ij})^2$.

For the general case when $\A_1\ne\A_2$, the situation is more complicated,
since it is non-trivial to construct a difference map 
which is free of sky signal.
A technique involving signal-to-noise 
eigenmode analysis has been derived for this case \cite{comparing},
but it is unfortunately rather complicated and cumbersome to implement.
Below we present a simpler method that eliminates need for this by
reducing the general problem to the simple case $\A_1=\A_2=\I$.

\section{Deconvolving maps}
\label{DeconvSec}

In this section, we present a method for 
inverting \eq{vEq}, \ie, for undoing the convolution with
beam and scanning effects given by the $\A$-matrices.

\subsubsection{Why is it useful?}

As we will see, this simplifies calculations by eliminating
all $\A$-matrices, encoding the corresponding complications 
and correlations
in the noise covariance matrices.

When comparing or combining two maps, it is generally undesirable to smooth
the higher resolution one down to the lower resolution of the other,
since this destroys information. Moreover, this tends to cause numerical
instabilities by making the smoothed noise covariance matrix poorly conditioned. 
We will see that, surprisingly, deconvolution can be better conditioned than
convolution/smoothing.

This deconvolution (elimination of $\A$-matrices) 
is useful not only for comparing
data sets as mentioned above, but for combining them as well.
The complication stems from the fact that the sky is sampled
by only a finite number of pixels, so to avoid problems with undersampling,
$\x$ must be the ``true sky'' beam-smoothed map with some finite
angular resolution.
If we do not deconvolve, but use equations\eqn{simEq} and\eqn{simSig} to combine two data
sets with different angular resolutions, say $\theta_1=0.89\arcdeg$
and $\theta_2=0.68\arcdeg$ for the QMAP Ka- and Q-band maps, respectively, 
it is natural to define $\x$ to have the higher 
($0.68\arcdeg$) resolution, setting $\A_2=\I$ and letting 
$\A_1$ incorporate the extra smoothing 
$\Delta\theta=\sqrt{\theta_1^2-\theta_2^2}$ in the lower resolution map.
The resulting map will now contain two kinds of pixels:
ones with resolution $\theta_2$ in the region covered only by
map 2 and with the higher resolution $\theta_1$ elsewhere.
If we need to combine this with a third map, the relevant smoothing
scale unfortunately becomes undefined near the boundary between the
two resolutions. Deconvolution eliminates all these problems.

\subsubsection{How does it work?}

In the generic case, deconvolution is strictly speaking impossible, 
since the matrix $\A$ is not 
invertible\footnote{Specifically, the problem is that $\A$ generically 
has a non-zero null space, \ie, that there are non-zero vectors $\x$ such that
$\A\x=\bzero$.} and certain pieces of 
information about $\x$ are simply not present in $\y$.
It is common practice to find approximate solutions to such under-determined
problems using singular value decomposition or other techniques, but
our goal here is different. We wish to compute a vector $\xt$
that can be analyzed as a true sky map. Specifically, we want analysis
of $(\xt,\Sig)$ to give exactly the same results
as analysis of $(\y,\N,\A)$ for all cosmological applications, 
say Wiener filtering or power spectrum estimation.

Our basic idea is to accept that certain modes in the map $\xt$ cannot 
be recovered, and to record this information in the noise covariance
matrix $\Sig$ by assigning a huge variance to these modes.
Any subsequent analysis (say Wiener filtering or power spectrum estimation)
will then automatically assign essentially zero weight to these modes.

In practice, we find it conceptually useful to imagine combining our data
$\y_1$ with a ``virtual map'' $\y_2$ that is so noisy that it contains
essentially no information, yet has 
the angular resolution $\theta$ that we wish to deconvolve down to,
\ie, $\A_2=\I$.
Although, as we will see, this virtual map never enters the calculations 
in practice, it is a useful notion for intuitively understanding what the
deconvolution technique does.
Specifically, let us take the noise in the virtual map to be uncorrelated,
with noise covariance matrix $\N_2=\sigma^2\I$ for some very large noise level
$\sigma$. Equations\eqn{simEq} and \eqn{simSig} give
\beqa{newsimEq}
\xt&=&\Sig\A_1^t\N_1^{-1}\y_1 + \sigma^{-2}\Sig\y_2,\\
\label{newsimSig}
\Sig&=&\left[\A_1^t\N_1^{-1}\A_1 + \sigma ^{-2}\I\right]^{-1}.
\eeqa
In the limit $\sigma\mapsto\infty$, 
$\xt$ will clearly become independent of the virtual temperature map $\y_2$ 
except for the ``junk modes'' which have infinite variance according 
to $\Sig$. For convenience, we therefore set $\y_2=\bzero$ in practice.\footnote{An 
alternative approach would be to set $\y_2$ equal to a 
Monte-Carlo generated map of independent Gaussian random variables with 
standard deviation $\sigma$
--- although this results in different 
numerical values in $\xt$, it will of course not change the results of any
subsequent cosmological analysis of $\xt$, since only the ``junk modes''
are different.
}

This deconvolution method has exactly the property we want as long as $\sigma$
is orders of magnitude larger than the pixel signal due to CMB.
If we were to choose $\sigma$ to be too small, then the virtual map 
would contribute a non-negligible amount of information and bias the
results. If we were to choose $\sigma$ to be too large, however, 
the matrix $\Sig$ would contain some enormous eigenvalues 
(since $\A_1^t\N_1\A_1$ is typically not invertible)
and be poorly conditioned, which could cause numerical problems in 
subsequent analysis. We performed a series of numerical tests to assess these
problems, and found that with $n\simlt 10^4$ pixels and double precision 
arithmetic, $\sigma=10^4\mu$K was a good compromise that 
produced neither of these two problems. We will therefore use
this choice throughout the present paper. 

\subsubsection{Tests}

As a first test of the method, we deconvolve (or ``unsmooth'')
a map $\y_1$ with resolution $\theta_1$ into a map with resolution 
$\theta_2$ $(\theta_2<\theta_1)$.
For this case, 
\beq{Amat}
\left(\A_1\right)_{ij} = 
{1\over 2\pi\Delta\theta^2}
e^{-{\theta_{ij}^2\over 2\Delta\theta^2}},
\eeq
where $\theta_{ij}=\cos^{-1}\left(\rh_i\cdot\rh_j\right)$ 
is the angular separation between pixels $i$ and $j$, and
$\Delta\theta\equiv\sqrt{\theta_1^2-\theta_2^2}$ is the extra smoothing to be undone. 
Specifically, we unsmooth 
the QMAP Ka-band data to obtain the same resolution as the Q-band data has,
from $0.89^\circ$ to $0.68^\circ$. 
We then Wiener filtered both the original and unsmoothed versions of
the map, obtaining virtually identical results.

As a second test, we deconvolve the Saskatoon data $\y$ into a map $\xt$
using the full (and rather complicated) $\A$-matrix described in
\cite{saskmap}. We then Wiener-filter $\xt$ and obtain a map virtually
identical to the one that was computed in \cite{saskmap} --- the latter was 
computed with a completely different method which circumvented the map step
altogether. The details of these maps have already been presented above.

Both of these tests thus confirm what we expect theoretically: that
the deconvolved map $(\xt,\Sig)$ contains exactly the same 
information about the true sky as the input data ($\y,\N,\A$), 
no more and no less.





\begin{references}   %



\bibitem{Gawiser00}
\rf\nn Gawiser E\dualand\nn Silk J;2000;Phys. Rept.;333-334;245-267
       
\bibitem{deBernardis00}
\rf\nn {de Bernardis} P {\etal};2000;Nature;404;955

\bibitem{Hanany00}
\rf\nn Hanany S {\etal};2000;ApJL;545;L5

\bibitem{Lange00}
\rfprep\nnn Lange A E {\etal};2000;astro-ph/0005004

\bibitem{boompa}
\rf\nn Tegmark M\dualand\nn Zaldarriaga M;2000;Phys. Rev. Lett.;85;2240    

\bibitem{Bambi00}
\rf\nn Balbi A {\etal};2000;ApJL;545;L1

\bibitem{Bridle00}
\rfprep\nnn Bridle S L {\etal};2000;astro-ph/0006170

\bibitem{observables}
\rn\nn Hu W, \nn Fukugita M, \nn Zaldarriaga M\multiand\nn Tegmark
M;astro-ph/0006436; {\it ApJ}, in press.

\bibitem{Jaffe00}
\rfprep\nn Jaffe A {\etal};2000;astro-ph/0007333

\bibitem{Kinney00}
\rfprep\nn Kinney W, \nn Melchiorri A\multiand\nn Riotto A;2000;astro-ph/0007375

\bibitem{concordance}
\rf\nn Tegmark M, \nn Zaldarriaga M\multiand\nnnn Hamilton A J S;2001;Phys. Rev. D;63;043007

\bibitem{Scott95}
\rf\nn Scott D, \nn Silk J\multiand\nn White M;1995;Science;268;829  
       
\bibitem{Lineweaver98}
\rf\nnn Lineweaver C H;1998;ApJL;505;L69

\bibitem{9par}
\rf\nn Tegmark M;1999;ApJL;514;L69
       	       
\bibitem{Dodel99}
\rf\nn Dodelson S\dualand\nn Knox L;2000;Phys. Rev. Lett;84;3523
       
\bibitem{10par}
\rf\nn Tegmark M\dualand\nn Zaldarriaga M;2000;ApJ;544;30T




\bibitem{Bennett96}
\rf\nnn Bennett C L \etal;1996;ApJ;464;L1


\bibitem{Ganga93}
\rf\nn Ganga K, \nn Cheng E, \nn Meyer S\multiand\nn Page L;1993;ApJL;410;L57

\bibitem{Lineweaver95}
\rf\nnn Lineweaver C H \etal;1995;ApJ;448;482

\bibitem{Knox98}
\rfprep\nn Knox L, \nnn Bond J R, \nnn Jaffe A H, 
\nn Segal M\multiand\nn Charbonneau D;1998;astro-ph/9803272

\bibitem{Netterfield97}
\rn\nnn Netterfield C B, \nnn Devlin M J, \nn Jarosik N, 
\nnn Page L A\multiand\nnn Wollack E J; ApJ, {\bf 474}, 47 (1997)
     
\bibitem{Ruhl95}
\rf\nnn Ruhl J E {\etal};1995;ApJL;453;L1

\bibitem{saskmap}
\rf\nn Tegmark M \etal;1996a;ApJL;474;L77

\bibitem{Inman97}
\rf\nnn Inman C A \etal;1997;ApJL;478;L1

\bibitem{qmap1}
\rf\nn Devlin M, \nn{de Oliveira-Costa} A, \nn Herbig T,
\nnn Miller A D, \nnn Netterfield C B, 
\nnn Page L A\multiand\nn Tegmark M;1998;ApJL;509;L77

\bibitem{qmap2}
\rf\nn Herbig T {\etal};1998;ApJL;509;L73

\bibitem{qmap3}
\rf\nn{de Oliveira-Costa} A, \nn Devlin M, \nn Herbig T,
\nnn Miller A D, \nnn Netterfield C B, 
\nnn Page L A\multiand\nn Tegmark M;1998;ApJL;509;L77

\bibitem{comparing}
\rf\nn Tegmark M;1999;ApJ;519;513

\bibitem{r}
\rf\nn Tegmark M\dualand\nnn Bromley B C;1999;ApJL;518;L69

\bibitem{EfstathiouNullbuster}
\rf\nn Seaborne M {\etal};1999;MNRAS;309;89

\bibitem{Netterfield}
\rf\nnn Netterfield C B {\etal};1995;ApJ;445;L69
     
\bibitem{saskforeg}
\rf\nn {de Oliveira-Costa} A, \nn Kogut A, \nnn Devlin M J, 
\nnn Netterfield C B, \nnn Page L A, \nnn Wollack E J;1997;ApJ;482;L17 

\bibitem{Smoot}
\rf\nnn Smoot G F {\etal};1992;ApJ;396;L1

\bibitem{Bennett}
\rf\nnn Bennett C L {\etal};1996;ApJ;464;L1

\bibitem{foregpars}
\rf\nn Tegmark M, \nnn Eisenstein D J, 
\nn Hu W\multiand\nn {de Oliveira-Costa} A;2000;ApJ;530;133 

\bibitem{qmapforeg}
\rf\nn {de Oliveira-Costa} A, \nn Tegmark M, \nnn Devlin M J, \nnn Haffner L M Haffner, 
\nn Herbig T, \nnn Miller A D, \nnn Page L A, \nnn Reynolds R J, 
\nnn Tufte S L;2000;ApJL;542;L5

\bibitem{gaussianity}
\rf\nnn Bromley B C\dualand\nn Tegmark M;1999;ApJL;524;L79

\end{references}
\end{document}